# Co-evolution of Functional Brain Network at Multiple Scales during Early Infancy

Xuyun Wen[a,b], Liming Hsu[b], Weili Lin[b], Han Zhang[b,*], Dinggang Shen[b,c,*]

[a] College of Computer Science and Technology, Nanjing University of Aeronautics and Astronautics, China
[b] Department of Radiology and BRIC, University of North Carolina at Chapel Hill, Chapel Hill, NC, USA
[d] Department of Brain and Cognitive Engineering, Korea University, Seoul 02841, Republic of Korea

**Abstract.** The human brains are organized into hierarchically modular networks facilitating efficient and stable information processing and supporting diverse cognitive processes during the course of development. While the remarkable reconfiguration of functional brain network has been firmly established in early life, all these studies investigated the network development from a "single-scale" perspective, which ignore the richness engendered by its hierarchical nature. To fill this gap, this paper leveraged a longitudinal infant resting-state functional magnetic resonance imaging dataset from birth to 2 years of age, and proposed an advanced methodological framework to delineate the multi-scale reconfiguration of functional brain network during early development. Our proposed framework is consist of two parts. The first part developed a novel two-step multi-scale module detection method that could uncover efficient and consistent modular structure for longitudinal dataset from multiple scales in a completely data-driven manner. The second part designed a systematic approach that employed the linear mixed-effect model to four global and nodal module-related metrics to delineate scale-specific age-related changes of network organization. By applying our proposed methodological framework on the collected longitudinal infant dataset, we provided the first evidence that, in the first 2 years of life, the brain functional network is co-evolved at different scales, where each scale displays the unique reconfiguration pattern in terms of modular organization.

## 1    Introduction

Early infancy is increasingly recognized as one of the most pivotal period of life with dramatic changes in brain function that largely shape the subsequent cognitive abilities [1]. Many recent studies were thus focusing on uncovering the developmental patterns of functional brains during this period by using resting-state functional magnetic resonance imaging (rs-fMRI) with diverse methodological approaches [2-5].

Among various methods, detecting modular organization of brain network attracted much attention because it could simultaneously characterize the development of 1) meso-scale brain network by tracking the formation and dissolution of modules across age; and 2) network segregation and integration by delineating growth trajectories of intra- and inter-modular functional connections [3, 4]. Although there are studies have adopted this approach to characterize early brain development [2, 3], all of them investigated network's modular organization from a "single-scale" perspective, which ignore the richness engendered by its hierarchical nature.

Abundant evidence has indicated that the brain's modular organization has a hierarchical structure where large-scale modules can be subdivided into smaller modules at finer scales [6, 7]. Such a network property is helpful for facilitating efficiency, robustness, adaptivity, and evolvability of network functions. Each scale might display the specific modular reconfiguration across age that provides the unique information for uncovering the pattern of brain network development [8]. However, to date, the studies simultaneously focusing on multiple scales for modular development are very scare, which is due to the lack of efficient multi-scale module detection method for longitudinal dataset and systematical approach for delineating scale-specific age-related changes of modular organization.

To fill above gaps, we collected a large longitudinal infant rs-fMRI dataset acquired at around neonate, 1 year old, and 2 years old, and proposed an advanced methodological framework to characterize multi-scale reconfiguration of functional brain network during early development. Our proposed framework is consist of two parts. The first part developed a two-step module detection method for the longitudinal dataset that could 1) uncover efficient multi-scale modular structure of the network in a completely data-driven manner; and 2) ensure the consistence of modular scales across age. The second part designed a systematic approach that employed the linear mixed-effect model to four modular-related metrics to delineate scale-specific developmental trajectories of network segregation and integration across age. By applying this framework on our collected longitudinal infant dataset, we, for the first time, investigated the early development of brain network from multiple scales, and found that, in the first 2 years of life, the brain network is co-evolved at different scales, where each scale displays the unique and specific modular reorganization.

## 2    Dataset and Methods

### 2.1    Dataset

Images were obtained from 48 typically developing term-born infants. 90 longitudinal rs-fMRI scans distributed at neonatal ($N = 33$), ~1-year-old ($N = 35$), and ~2-year-old ($N = 22$) were collected with a Siemens 3T MRI scanner by using a T2-weighted EPI sequence with parameters: TR = 2s, TE = 32ms, 33 slices, voxel size = $4 \times 4 \times 4 mm^3$, and the total number of volumes = 150.

The infant rs-fMRI data was preprocessed with FSL software including the following steps: 1) discarding the first 10 volumes; 2) slice-timing correction; 3) head motion correction; 4) spatial smoothing; 5) low-pass temporal filtering (<0.08 Hz); 6) mean signal removal (including white matter, cerebrospinal fluid (CSF), whole-brain averaged signal and six motion parameters); 7) data scrubbing to control the global measure of signal changes ($< 0.5\%$) and frame-wise displacement ($< 0.5$ mm) [9]; and 8) spatial registration to register each rs-fMRI data from its native space to the Montreal Neurological Institutes (MNI) space by using an infant-dedicated algorithm.

After data preprocessing, we constructed the functional brain network for each subject at each age. For each scan, we firstly parcellated the whole brain into 180 brain regions (exclude cerebellum) by using atlas provided in [10], and constructed the corresponding functional network based on regions' averaged time courses. The regional pair-wise functional connectivity (FC) was measured by a $z$ value, i.e., Pearson's correlation after Fisher-z transformation. We then set all negative values in each FC matrix to zeros and averaged all individual FC matrices of each age group as the age-specific group-level functional network. After that, we conducted two-tailed one-sample t-tests on all 16,110 ($180 \times 179/2$) pairwise FCs across all subjects of each age group and only kept the connections passing the test ($p<0.05$ after Bonferroni-correction) to reduce the effect of external noise on the network construction [11]. Hence, we totally constructed 90 individual-level weighted FC matrices (90 scans) and 3 averaged group-level weighted FC matrices (3 age groups).

## 2.2 Two-step Multi-scale Module Detection Method for Longitudinal Dataset

Most of longitudinal rs-fMRI data is collected at several fixed time points, as our infant dataset mainly distributed at three age groups, i.e., neonates, 1-year-olds, and 2-year-olds. Thus, the multi-scale module detection for this type of dataset could be converted as a multi-scale multi-group module detection problem. Based on this motivation, our study proposed a two-step multi-scale module detection method for longitudinal dataset as flowcharted in Fig. 1, where the first step developed a multi-scale module detection algorithm for efficiently uncovering modular structure of the network at each of time points (or age groups) from different spatial scales, and the second step designed a new index (*BSS*) and used it as the measurement for determining modular-scale correspondence across age. The detailed description of each step was described as below.

**Step 1: Partition-similarity-based Multi-scale Module Detection method.** The most widely adopted method for module detection is through maximizing modularity ($Q$) measurement [12] defined as

$$Q = \frac{1}{2m} \sum_{ij} \left( a_{ij} - \gamma \frac{k_i k_j}{2m} \right) \delta(i,j) \tag{1}$$

where $a_{ij}$ represents link strength between nodes $i$ and $j$, and $k_i$ indicates the degree of node $i$. $m$ is the total link strength of the network. The Kronecker delta function, $\delta(i,j)$ = 1, when nodes $i$ and $j$ belong to the same module; otherwise, $\delta(i,j) = 0$. The parameter $\gamma$ controls the resolution of detected modules that a smaller value corresponds to a coarser scale, and a larger one corresponds to a finer scale. However, in the absence of prior hierarchical information of modules, this traditional method could not provide efficient multi-scale module detection results because it is difficult to select the proper *γ*s that cover all modular scales while avoiding redundancy. To address this issue, our study proposed to construct the partition similarity matrix (*PSM*) and introduce graph clustering for partition division, and thus developed a *p*artition-*s*imilarity-based **M**ulti-scale **M**odule **D**etection method (i.e., *ps*-MSMD) to efficiently uncover multi-scale modular structure of the network in a completely data-drive manner.

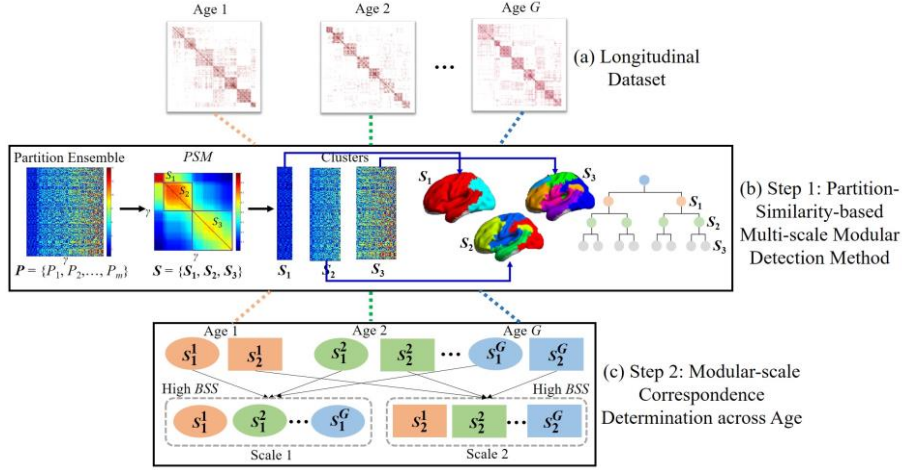

**Fig. 1.** Illustration of the proposed two-step multi-scale module detection method. With multiple networks from longitudinal dataset input, it outputs multi-scale module detection results.

As shown in Fig. 1(b), *ps*-MSMD is consist of three steps. *First*, we varied parameter $\gamma$ in Eq. 1 to generate a partition ensemble with modular structure from low resolutions to high resolutions represented as $\boldsymbol{P} = \{P_1, P_2, \ldots, P_m\}$, where each $P_i$ is a network partition obtained under $\gamma_i$. *Second*, we constructed *PSM*, in which each element $PS_{ij}$ represents the similarity between $P_i$ and $P_j$ measured by Jaccard's index (ranges from [0, 1], and a larger value indicates higher similarity), and conducted graph clustering [13] on *PSM* to divide partition ensemble $\boldsymbol{P}$ into $n$ clusters represented as $\boldsymbol{S} = \{\boldsymbol{S_1}, \boldsymbol{S_2}, \ldots, \boldsymbol{S_n}\}$. Each $\boldsymbol{S_i} = \{P_i^1, P_i^2, \ldots, P_i^q\}$ contains $q$ different partitions with similar modular structure, thus representing one modular scale. In this way, we determined the number of modular scales (i.e., the number of clusters of *PSM*) and assigned each partition in $\boldsymbol{P}$ into one scale. *Third*, for each scale $i$, we selected the partition that is most similar to others in $\boldsymbol{S_i}$ as the representative, and constructed the corresponding modular partition by using consensus clustering approach [14].

**Step 2: Modular-scale Correspondence Determination across Age.** As shown in Fig. 1(c) that each age group displays more than one modular partitions when detected with *ps*-MSMD, it is essential to determine the correspondence of modular scales across ages to ensure the investigation of modular development in a consistent man-ner. Considering the modular structure across age at the same scale should be more similar than those coming from different scales, we proposed to calculate the across-age across-scale modular similarity and assigned those partition ensembles at different age groups with the highest similarity into one scale. In light of this, we defined a new evaluation index, i.e., *BSS*, as below:

$$BSS(\boldsymbol{S_i^g}, \boldsymbol{S_j^h}) = \frac{1}{|\boldsymbol{S_i^g}| \times |\boldsymbol{S_j^h}|} \sum_{P_s \in S_i^g} \sum_{P_t \in S_j^h} \mathrm{J}(P_s, P_t), g \neq h \qquad (2)$$

where $\boldsymbol{S_i^g}$ is the partition ensemble of age group $g$ at $i$-th modular scale, and $|\boldsymbol{S_i^g}|$ is the number of partitions in $\boldsymbol{S_i^g}$. J represents Jaccard's index for measuring partition similarity. Thus, *BSS* computed the averaged similarity of partition ensemble between two scales from different age groups. It ranges from [0, 1], and a larger *BSS* represents a higher between-scale similarity. Notably, different age groups might detect different numbers of modular scales, and our study only focused on those scales emerging at all age groups.

### 2.3 Developmental Trajectories of Brain Network Organization

After determining the modular structure of each age group at each scale, we warped the obtained network partition to each individual's FC network and then adopted two global modular-related metrics (i.e., mean intra- and inter-modular FC, m$FC_w$ and m$FC_b$) and two nodal modular-related metrics (i.e., within-modular degree (*WD*) [15] and participation coefficient (*PC*) [16]) to characterize its modular segregation and integration. m$FC_w$ (m$FC_b$) was defined as the averaged FC strength across all within-modular (between-modular) connections. *WD* and *PC* of node $i$ were respectively calculated with

$$WD_i = \frac{K_{p_c}^i - \overline{K_{p_c}}}{\delta_{K_{p_c}}} \qquad (3)$$

$$PC_i = 1 - \sum_{c=1}^{C} \left( \frac{K_{p_c}^i}{k_i} \right) \qquad (4)$$

where $K_{p_c}^i$ is the total FC strength between node $i$ and the other nodes in module $p_c$. $\overline{K_{p_c}}$ and $\delta_{K_{p_c}}$ respectively represent the average and standard deviation of $K_{p_c}^i$ across all nodes in module $p_c$. $k_i$ is the sum of FC strength connecting node $i$. Among four metrics, m$FC_w$ and *WD* evaluated intra-modular FC that measures network segregation, and m$FC_b$ and *PC* evaluated inter-modular FC that measures network integration.

After obtaining m$FC_w$, m$FC_b$, and nodal *WD* and *PC* of all subjects at each age, we adopted linear mixed-effect regression (LMER) model to delineate their developmental trajectories across age. For each measurement, two LMER models were built, one of which uses age as fixed-effect variable (linear model), and the other uses log(age) as fixed-effect variable (log-linear model). Random intercept and subject were included as random effects. The Akaike information criterion (AIC) was adopted for model selection. For all models, the significance level was set to $p < 0.05$ after false discovery false (FDR) correction.

## 3    Results

### 3.1    Method Evaluation of *ps*-MSMD

This section evaluated the effectiveness and robustness of *ps*-MSMD on a synthetic network with two-level modular structures and a public test-retest adult rs-fMRI dataset (http://dx.doi.org/10.15387/fcp_indi.corr.hnu1). The synthetic network was generated with Lancichinetti-Fortunato-Radicchi model [17] with following parameters: $N = 200$, $k = 5$, $k_{max} = 10$, $\mu_1 = 0.1$, $\mu_2 = 0.1$, $c_{min} = 10$, $c_{max} = 20$, $C_{min} = 50$, $C_{max} = 100$. The detailed data information, image preprocessing and functional network construction for adult rs-fMRI data were referred to study [18]. For synthetic network, *ps*-MSMD successfully detected two-level modular partitions and both of them displayed the good precision as shown in Fig. 2a (coarser scale: Jaccard index = 0.96; finer scale: Jaccard index = 0.89), indicating our proposed method is efficient for multi-scale module detection. For adult's FC network, *ps*-MSMD detected two hierarchies of modular structure at both test and retest data, one of which includes four modules (scale 1, left figures in Fig. 2b), the other has eight modules (scale 2, right figures in Fig. 2b). The highly similar result between test and retest data indicates our proposed method is robust for multi-scale module detection.

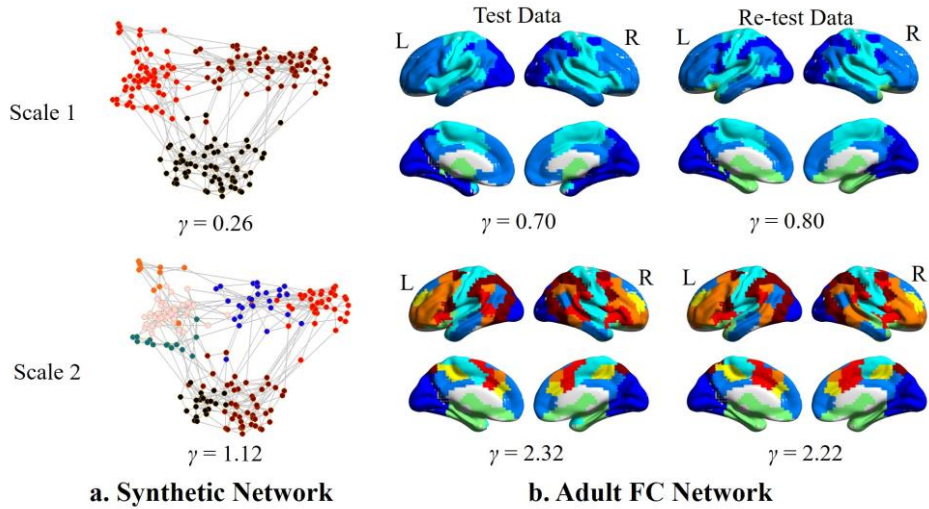

**a. Synthetic Network**          **b. Adult FC Network**

**Fig. 2.** Evaluation of *ps*-MSMD on (a) a synthetic network and (b) a public test-retest adult rs-fMRI dataset. Different modules were coded with different colors. The selected $\gamma$ for generating each partition was listed below each sub-figure.

### 3.2 Multi-scale Module Detection on Infant Dataset

In this section, we applied the proposed two-step multi-scale module detection method to the longitudinal infant dataset to uncover how modular structure develops in the first 2 years of age at different scales. By using *ps*-MSMD, we detected two partition clusters for neonates (labelled as $S_1^1$ and $S_2^1$) and 1-year-olds (labelled as $S_1^2$ and $S_2^2$), and three clusters for 2-year-olds (labelled as $S_1^3$, $S_2^3$ and $S_3^3$). To determine modular-scale correspondence across age, we measured across-age across-scale modular similarity with *BSS*, and assigned those partition ensembles with the highest *BSS* into one scale. Thus, $S_1^1$, $S_1^2$ and $S_1^3$ composed scale 1 (i.e., coarser scale), and $S_2^1$, $S_2^2$ and $S_2^3$ formed scale 2 (i.e., finer scale) as shown in Fig. 3a. For each detected scale at each age, we constructed the modular partition based on its corresponding partition ensemble and visualized it on brain surface in Fig. 3b.

From Fig. 3b, we observed that, in the first 2 years of life, the spatial distribution of modules at both coarser and finer scales was developed from local, anatomically proximal groups to distributed functional systems, during which diverse functional sub-networks were evolved to be more and more adult-like. However, the time period with the most dramatic modular changes was different between two scales. Specifically, at the coarser scale, the modular structure displayed more dramatic changes from 1 year old to 2 years old, during which the number of functional modules was increased from 4 to 6. However, at the finer scale, the larger reorganization of modular structure occurred in the first year after birth. During this period, the prototypes of most high-order functional systems were well shaped, including default mode network, frontal-parietal network, and ventral attention network.

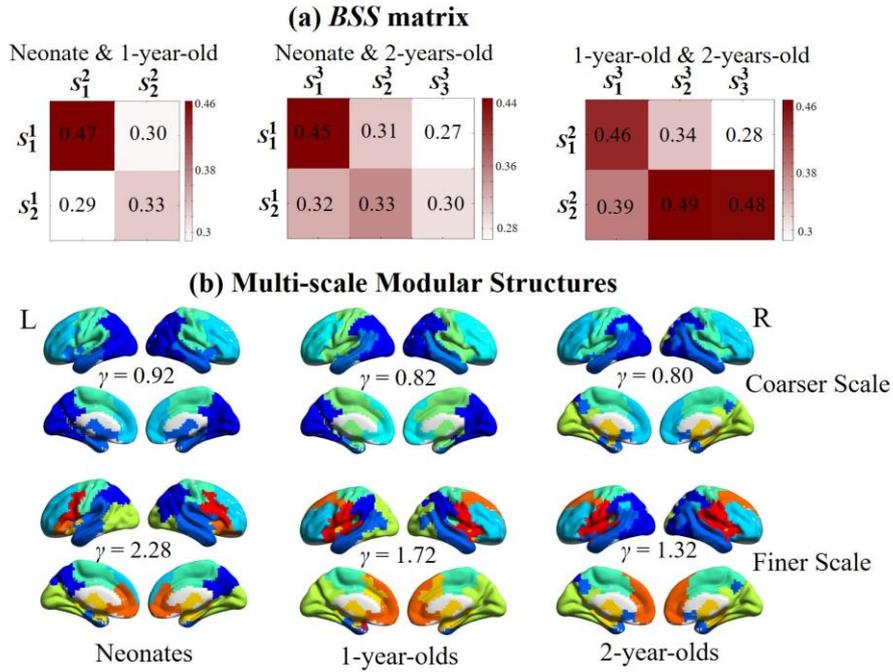

**Fig. 3.** Multi-scale module detection on the longitudinal infant dataset by using our proposed method. (a) *BSS* matrix between each pair of age groups calculated on the detected multi-scale modular partitions with *ps*-MSMD; (b) Visualization of module detection results on infant dataset at different scales, where different colors coded different modules. The selected $\gamma$ for generating each partition was listed inside each sub-figure.

### 3.3 Development of Brain Network Organization

Based on developing modular structure of each scale, we further characterized scale-specific age-related brain network reconfiguration by delineating developmental trajectories of four modular-related metrics, i.e., m$FC_\mathrm{w}$, m$FC_\mathrm{b}$, *WD* and *PC*. All results on four metrics suggest that, *from birth to 2 years of age, the brain network displays co-evolution at all scales, while each scale displays the unique modular reconfiguration pattern*. Specifically, at the coarser scale, both network segregation and integration were enhanced in the first 2 years of life, which are respectively supported by the increasing developmental trajectory of m$FC_\mathrm{w}$ (Fig. 4a), and the result of much more brain regions displaying increased *PC* than those in decrease (Fig. 4d). However, at the finer scale, only the network segregation was found to be increased during early development indicated by the number of regions with increasing *WD* larger than those in decrease (Fig. 4c). For the network integration, our current results from m$FC_\mathrm{b}$ (Fig. 4b) and *PC* (Fig. 4d) could not support its enhancement in the first 2 years of life.

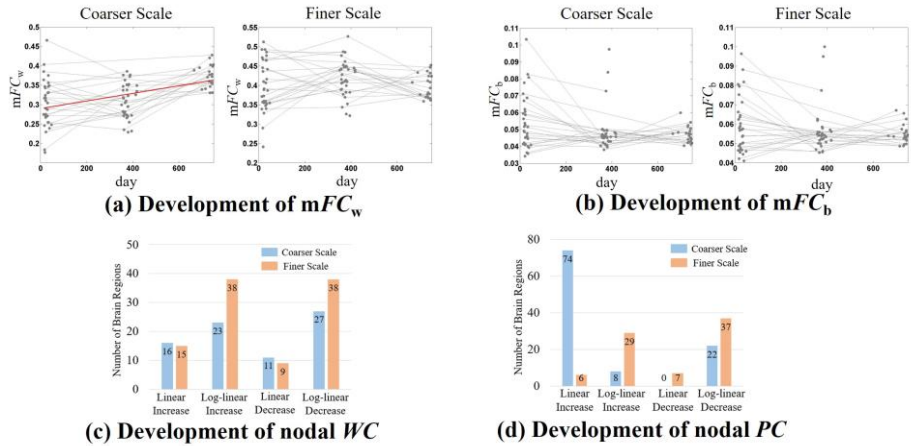

**Fig. 4.** Age-related scale-specific changes of brain functional network measured by (a) m$FC_\mathrm{w}$, (b) m$FC_\mathrm{b}$, (c) nodal *WD* and (d) nodal *PC* in the first 2 years of life. In Figs. (a) and (b), only statistically significant increasing curves were fitted with red lines.

## 4    Conclusions

To our knowledge, this paper provides the first multi-scale developmental report of functional brain network in the first 2 years of life. To this end, we firstly proposed an advanced methodological framework to delineate multi-scale reconfiguration of brain network during development, which includes two parts. The first part developed a two-step multi-scale module detection method for longitudinal dataset to uncover efficient and consistent modular structure of the network from multiple scales in a completely data-driven manner. The second part proposed a systematical approach that employed the linear mixed-effect model to four modular-related metrics to characterize scale-specific age-related changes of network segregation and integration. We then applied the proposed framework to a large longitudinal infant resting-state functional magnetic resonance imaging dataset, and found the co-evolution of brain functional network from birth to 2 years of age.